\documentclass[amsmath,amssymb,floatfix,reprint,twocolumn,superscriptaddress]{revtex4-1}

\usepackage{graphicx}
\usepackage{epsfig}
\usepackage{hyperref}
\usepackage{xcolor}
\usepackage{bm}

\begin{document}

\title{Direct measurement of a remnant Fermi surface in $\mathbf{SmB_{6}}$}

\author{Thomas E. Millichamp}
\affiliation{H.H. Wills Physics Laboratory, University of Bristol, Tyndall Avenue, Bristol, BS8 1TL, United Kingdom}
\affiliation{The Polesworth School, Dordon Road, Dordon, Tamworth, B78 1QT, United Kingdom}
\author{David Billington}
\email{billingtond1@cardiff.ac.uk}
\affiliation{School of Physics and Astronomy, Cardiff University, Queen's Building, The Parade, Cardiff, CF24 3AA, United Kingdom}
\affiliation{Japan Synchrotron Radiation Research Institute, SPring-8, Sayo 679-5198, Japan}
\author{Hannah C. Robarts}
\affiliation{H.H. Wills Physics Laboratory, University of Bristol, Tyndall Avenue, Bristol, BS8 1TL, United Kingdom}
\author{Jude Laverock}
\affiliation{H.H. Wills Physics Laboratory, University of Bristol, Tyndall Avenue, Bristol, BS8 1TL, United Kingdom}
\affiliation{School of Chemistry, University of Bristol, Cantock's Close, Bristol, BS8 1TS, United Kingdom}
\author{Daniel O'Neill}
\affiliation{Department of Physics, University of Warwick, Coventry, CV4 7AL, United Kingdom}
\author{Monica Ciomaga Hatnean}
\affiliation{Department of Physics, University of Warwick, Coventry, CV4 7AL, United Kingdom}
\affiliation{Materials Discovery Laboratory, Department of Materials, Swiss Federal Institute of Technology Zurich, CH-8093 Zurich, Switzerland}
\affiliation{Laboratory for Multiscale materials eXperiments, Paul Scherrer Institut, CH-5232 Villigen PSI, Switzerland}
\author{Geetha Balakrishnan}
\affiliation{Department of Physics, University of Warwick, Coventry, CV4 7AL, United Kingdom}
\author{Jonathan A. Duffy}
\affiliation{Department of Physics, University of Warwick, Coventry, CV4 7AL, United Kingdom}
\author{Jonathan W. Taylor}
\affiliation{DMSC - European Spallation Source, Universitetsparken 1, Copenhagen 2100, Denmark}
\author{Sean R. Giblin}
\affiliation{School of Physics and Astronomy, Cardiff University, Queen's Building, The Parade, Cardiff, CF24 3AA, United Kingdom}
\author{Stephen B. Dugdale}
\email{s.b.dugdale@bristol.ac.uk}
\affiliation{H.H. Wills Physics Laboratory, University of Bristol, Tyndall Avenue, Bristol, BS8 1TL, United Kingdom}

\date{\today}

\maketitle

{\bf
The quest to understand the nature of the electronic state in $\mathbf{SmB_{6}}$ has been challenging, perplexing and surprising researchers for over half a century.
In the theoretically predicted topological Kondo insulator $\mathbf{SmB_{6}}$ \cite{Dzero2010a}, the nature of the bulk electronic structure is not characterised unambiguously by quantum oscillations due to contrary interpretations \cite{Li2014a,Tan2015a}.
One simple definition of an electrical insulator is a material that lacks a Fermi surface and here we report the results of our investigation into its existence in $\mathbf{SmB_{6}}$ by Compton scattering \cite{Cooper85}.
Compton scattering measures occupied electron momentum states, is bulk sensitive due to the high energy of the incoming photons \cite{Dugdale2014a} and is also an ultra-fast probe
of the correlated many-body electron wavefunction \cite{Cooper1993a}.
Remarkably, direct evidence for a three-dimensional remnant Fermi surface is observed.
However, a further dichotomy is raised in that the full occupancy expected of a conventional metal is not reproduced.
Our observation of a remnant Fermi surface using a momentum-resolved probe provides significant new evidence that the paradigm of a bulk insulator is not robust for $\mathbf{SmB_{6}}$.
}

The temperature dependence of the resistivity in SmB$_{6}$
led to the characterisation of the material as a hybridised $f$-electron Kondo insulator at low temperature.
However, the recent focus on topologically nontrivial band structures has led to
the claim that SmB$_{6}$ is a topological Kondo insulator \cite{Dzero2010a}.
As with conventional topological insulators, there should be clear evidence
of surface states and, experimentally, a combination of both resistivity \cite{Kim2013a,Zhang2013a,Kim2014a}
and angle-resolved photoemission spectroscopy (ARPES) \cite{Xu2013a,Neupane2013a,Jiang2013a}
have suggested that such states are indeed present in SmB$_{6}$.
More recent measurements have further shown that the surface states are spin-polarised
with the spin locked to the crystal momentum \cite{Xu2014a}, as required for topologically protected states.
Conversely, the exact nature of the {\it bulk} electronic structure is still a somewhat open question \cite{Zhang2020}.
Two experiments utilising the de Haas-van Alphen (dHvA) technique have come to opposite conclusions as to the origin of the observed quantum oscillations.
One reports that the angular dependence of the oscillations is evidence of
2D electron orbits associated with the surface \cite{Li2014a}, while the other finds that a
3D Fermi surface is required to explain their results \cite{Tan2015a}.

Although the existence of a bulk Fermi surface may seem paradoxical for a topological Kondo insulator, there is the possibility that it could hide charge-neutral quasiparticles within the bulk \cite{Tan2015a,Chowdhury17}.
Investigations from a variety of techniques including muon spin relaxation
($\mu$SR) \cite{Biswas2014a,Biswas2017a},
AC conductivity \cite{Laurita2016a}, neutron scattering \cite{Alekseev1993,Fuhrman2015a} and specific heat \cite{Wakeham2016a,Hartstein17,Phelan2014a}
have already found evidence for bulk in-gap excitations.
These experiments do not directly probe the Fermi surface
and have led to differing interpretations of the experimental data within the topological paradigm \cite{Denlinger2016a}.
Further theoretical proposals to explain previous data include excitons \cite{Knolle2017a}, Majorana fermions \cite{Baskaran2015a} and Kondo breakdown \cite{Erten2016a}, with different implications for the putative Fermi surface.

\begin{figure*}[t!]
\centerline{\includegraphics[width=1.0\linewidth]{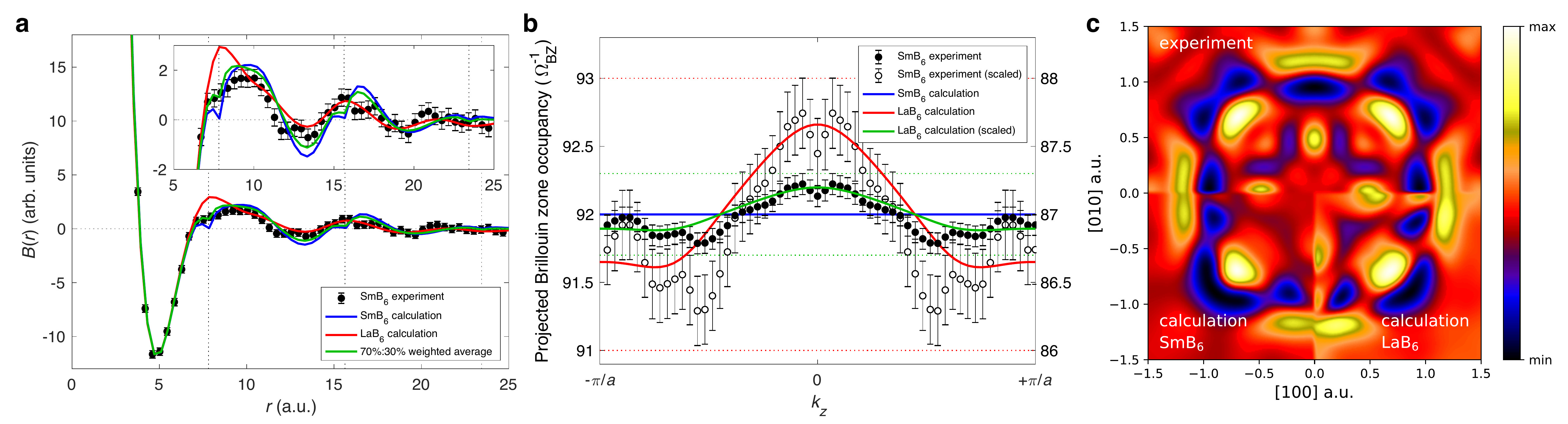}}
\caption{
{\bf Nature of the bulk electronic state in $\mathbf{SmB_{6}}$.}
{\bf a}, Experimental $B(r)$ for the $p_{z}\parallel[100]$ Compton profile of SmB$_{6}$ together with those calculated for SmB$_{6}$ (insulating), LaB$_{6}$ (metallic), and a $7:3$ weighted average of the $B(r)$ functions for SmB$_{6}$ and LaB$_{6}$.
Vertical dotted lines indicate integer multiples of the $[100]$ lattice constant, $a=7.815$~a.u., for which $B(r)$ must be zero for an insulator in a one-electron framework.
The inset shows a magnified portion of the main panel.
{\bf b}, Experimental $\rho_{\rm 1D}(k_{z})$ for the $p_{z}\parallel[100]$ Compton profile of SmB$_{6}$ together with those calculated for SmB$_{6}$ and LaB$_{6}$.
The horizontal dotted lines are $\rho_{\rm 1D}=(N_{e}\pm1)/\Omega_{\rm BZ}$ and $\rho_{\rm 1D}=(N_{e}\pm0.3)/\Omega_{\rm BZ}$, where $N_{e}=92$ and $87$ for SmB$_{6}$ (left axis) and LaB$_{6}$ (right axis), respectively, and here $\Omega_{\rm BZ}=2\pi/a$ is the $[100]$-line projected 1D Brillouin zone length.
Also shown are the experimental SmB$_{6}$ distribution scaled relative to $N_{e}/\Omega_{\rm BZ}$ to fit the LaB$_{6}$ calculation and {\it vice versa}.
{\bf c}, Experimental (upper half) radial anisotropy of the $(001)$-plane projected $\rho_{\rm 2D}(p_{x},p_{y})$ of SmB$_{6}$ alongside those calculated for SmB$_{6}$ (bottom left) and LaB$_{6}$ (bottom right) to show the resemblance of the anisotropies, indicative of ellipsoidal electron Fermi surface sheets.
The calculated distributions in {\bf b} and {\bf c} have been convoluted with 1D and 2D Gaussian functions, respectively, approximating the experimental momentum resolution.
Error bars in {\bf a} and {\bf b} indicate statistical errors of one standard deviation, demonstrating that the observed metallic signatures are robust.
}
\label{aniso}
\end{figure*}

High-resolution x-ray Compton scattering probes the electron momentum density, $\rho({\bm p})$,
which is the probability density of electrons in momentum space (${\bm p}$-space) \cite{Cooper85}.
The technique measures the so-called Compton profile, $J(p_{z})$, which is defined as the 1D projection
of $\rho({\bm p})$, resolved parallel to the scattering vector, $J(p_{z})=\iint\rho({\bm p})~\text{d}p_{x}\text{d}p_{y}$.
Compton scattering provides a unique probe of the correlated (many-body) ground-state electronic wavefunction via its underlying momentum distribution, yet is insensitive to lattice defects and atomic disorder.
Since only occupied electronic states can contribute to $\rho({\bm p})$, it also contains information about
the Fermi surface in a metal.
Being equally sensitive to all of the electrons means that $J(p_{z})$ can be straightforwardly normalised to the number of electrons per primitive cell, $N_{e}\equiv\int J(p_{z})~\text{d}p_{z}\equiv\iiint\rho({\bm p})~\text{d}^{3}{\bm p}$.
To provide a comparison with SmB$_{6}$, we refer to isostructural LaB$_{6}$ which is a conventional, monovalent Pauli paramagnetic metal whose Fermi surface is well-known \cite{Biasini1997a,Harrison:1993}.
Unlike SmB$_{6}$, whose anomalous properties have been associated with strong hybridisation between highly localised $4f$ states and delocalised $5d$ bands in the vicinity of the Fermi level ($E_{\rm F}$), the $4f$ states of LaB$_{6}$ lie a few eV above $E_{\rm F}$ so are unoccupied and do not contribute to its ground state properties.

While $J(p_{z})$ resides in momentum space, in real space (${\bm r}$-space) the reciprocal form factor (which is the autocorrelation of the ${\bm r}$-space wavefunction, see Methods), $B({\bm r})$,
can be obtained along a particular direction through ${\bm r}$-space by taking the 1D Fourier transform of a directional Compton profile, $B(r)=\int J(p)\text{e}^{\text{i}pr}~\text{d}p$.
In a one-electron framework, $B(r)$ for an insulator has to be zero at non-zero integer multiples of the lattice vector, ${\bm R}$.
This can be seen, for example, in the $B(r)$ functions for the weakly correlated (band) insulator, Si, and the strongly correlated (Mott) insulator, NiO (see Fig.~S1 and S2, respectively, in the Supplementary information).
The $B(r)$ function for the experimental $p_{z}\parallel[100]$ Compton profile of SmB$_{6}$ is shown in Fig.~\ref{aniso}{\bf a} together with those calculated for SmB$_{6}$ and LaB$_{6}$ which find insulating (gapped) and metallic (ungapped) ground states, respectively, in agreement with similar calculations \cite{antonov:02b,Hasegawa1977}.
In the experiment, it is evident that $B(r)$ is non-zero at integer multiples of the lattice constant, in contrast to what is predicted by the insulating SmB$_{6}$ calculation.
One advantage of transforming to real space is that the impact of finite experimental momentum resolution reduces to a straightforward scaling of $B(r)$ due to the usual Fourier convolution theorem, meaning that this cannot explain their absence.
The experimental $B(r)$ agrees with a metallic paradigm for SmB$_{6}$.

To further investigate the signatures of a metallic state, it is necessary to return to momentum space and back-fold (see Methods)
the ${\bm p}$-space distribution, $\rho({\bm p})$, into the first Brillouin zone to obtain the crystal momentum space (${\bm k}$-space) occupation density, $\rho({\bm k})$, because it is in ${\bm k}$-space that the Fermi surface exists.
In 3D, the obtained $\rho({\bm k})$ distribution, essentially, counts the number of electrons in occupied bands at every ${\bm k}$-point.
Thus, any variation in $\rho({\bm k})$ will be due the (de)population of the electron energy bands, $E_{i}({\bm k})$, as they cross the Fermi level, $E_{\rm F}$, at the Fermi wavevector, ${\bm k}_{\rm F}$, thus revealing the presence of a Fermi surface.
This back-folding procedure can be applied directly to a single directional Compton profile to obtain the 1D projected ${\bm k}$-space occupation density, $\rho_{\rm 1D}(k_{z})$.
Fig.~\ref{aniso}{\bf b} shows $\rho_{\rm 1D}(k_{z})$ determined directly from the experimental $p_{z}\parallel[100]$ Compton profile of SmB$_{6}$ together with those calculated for SmB$_{6}$ (insulating) and LaB$_{6}$ (metallic).
For SmB$_{6}$ the ${\bm k}$-space occupation density is flat, as expected for any insulator, because none of the bands cross $E_{\rm F}$ at any ${\bm k}$-point; any structure is a signature of partially occupied bands, {\it i.e.} a metal.
The experimental $\rho_{\rm 1D}(k_{z})$ for SmB$_{6}$ shows clear variation with $k_{z}$ and is further experimental evidence for it not being a conventional insulator.
Significantly, the intensity of the experimental $\rho_{\rm 1D}(k_{z})$ is suppressed to about $30\%$ of that calculated for metallic LaB$_{6}$.
We also note that the variation (near $k_{z}=\frac{\pi}{2a}$) is not identical to LaB$_{6}$.

\begin{figure*}[t!]
\centerline{\includegraphics[width=1.0\linewidth]{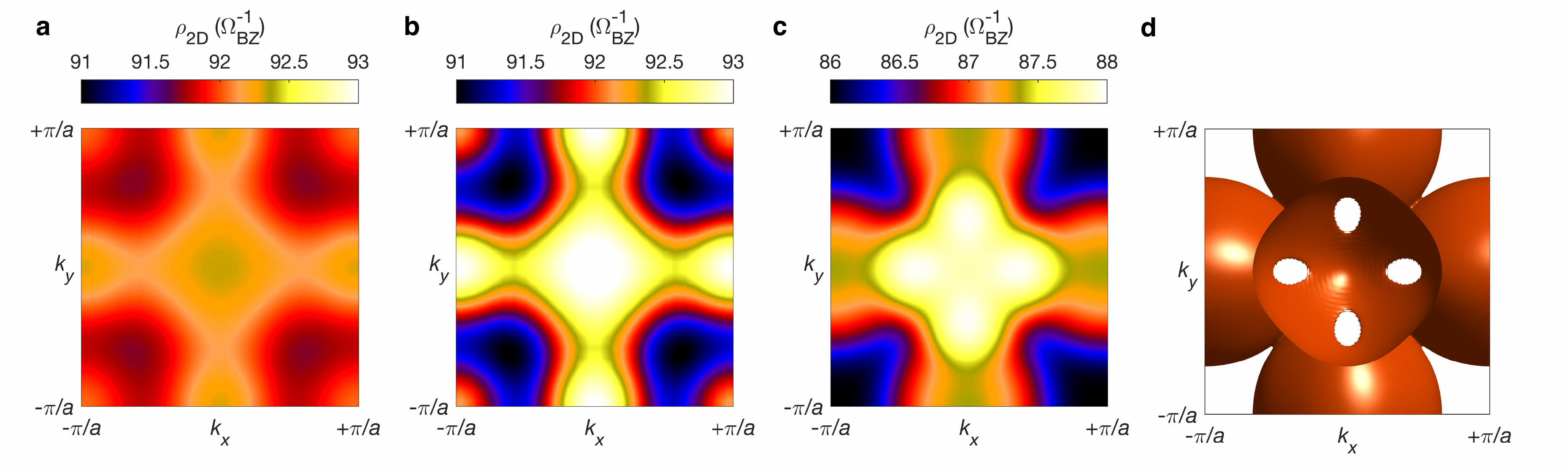}}
\caption{
{\bf Projected occupation density of $\mathbf{SmB_{6}}$ compared with $\mathbf{LaB_{6}}$.}
{\bf a}, Experimental $\rho_{\rm 2D}(k_{x},k_{y})$ for SmB$_{6}$ showing the startling variation across the
projected 2D Brillouin zone, indicating the existence of a Fermi surface.
{\bf b}, Same as {\bf a} but with the dynamic range scaled to give what would be expected for a monovalent metal.
{\bf c}, Calculated $\rho_{\rm 2D}(k_{x},k_{y})$ for LaB$_{6}$ convoluted with a 2D Gaussian approximating the
experimental momentum resolution.
{\bf d}, Calculated Fermi surface of LaB$_{6}$ given by the isoenergy surface $E_{j}({\bm k}_{\rm F})=E_{\rm F}$
for the single half-filled band which crosses $E_{\rm F}$.
Here, $\Omega_{\rm BZ}=(2\pi/a)^{2}$ is the $(001)$-plane projected 2D Brillouin zone area.
}
\label{OccDens}
\end{figure*}

Since each directional $J(p_{z})$ is a 1D projection of the 3D $\rho({\bm p})$, tomographic reconstruction
is required to recover any integrated-over (projected) components.
Here, Cormack's method, as developed by Kontrym-Sznajd \cite{Sznajd90}, was used to recover one component and provide the projection of $\rho({\bm p})$ onto the $(001)$-plane, $\rho_{\rm 2D}(p_{x},p_{y})=\int\rho({\bm p})~\text{d}p_{z}$, with $p_{z}\parallel[001]$.
The radial anisotropy (deviation from the angular average at constant momentum) of $\rho_{\rm 2D}(p_{x},p_{y})$, contains contributions from both the fully and any partially occupied bands.
Thus, in a metal, some features will be due to the presence of the Fermi surface.
The experimental radial anisotropy of $\rho_{\rm 2D}(p_{x},p_{y})$ for SmB$_{6}$ is shown in Fig.~\ref{aniso}{\bf c} together with those calculated for SmB$_{6}$ and LaB$_{6}$.
Superficially, the majority of the features are well-matched between the experiment and both the SmB$_{6}$ and LaB$_{6}$ calculations because the dominant contributions are from similar completely filled bands.
However, inspection of the experimental $\rho_{\rm 2D}(p_{x},p_{y})$ distribution reveals ellipsoidal features around the projected set of $\big\{\frac{\pi}{a},0\big\}$-points ($\pi/a\approx0.4$~a.u.) of the $(001)$-plane projected first Brillouin zone.
A similar feature can be seen in LaB$_{6}$ where it has been unambiguously associated with the $\big\{\frac{\pi}{a},0,0\big\}$-centred ellipsoids of its Fermi surface \cite{Biasini1997a,Harrison:1993} (see Fig.~S3 in the Supplementary information).
In the SmB$_{6}$ experiment, there is some additional intensity in the radial anisotropy located around the projected set of $\big\{\frac{\pi}{a},\frac{\pi}{a}\big\}$-points which is not seen in either the insulating SmB$_{6}$ or metallic LaB$_{6}$ calculations.

\begin{figure}[t!]
\centerline{\includegraphics[width=1.0\linewidth]{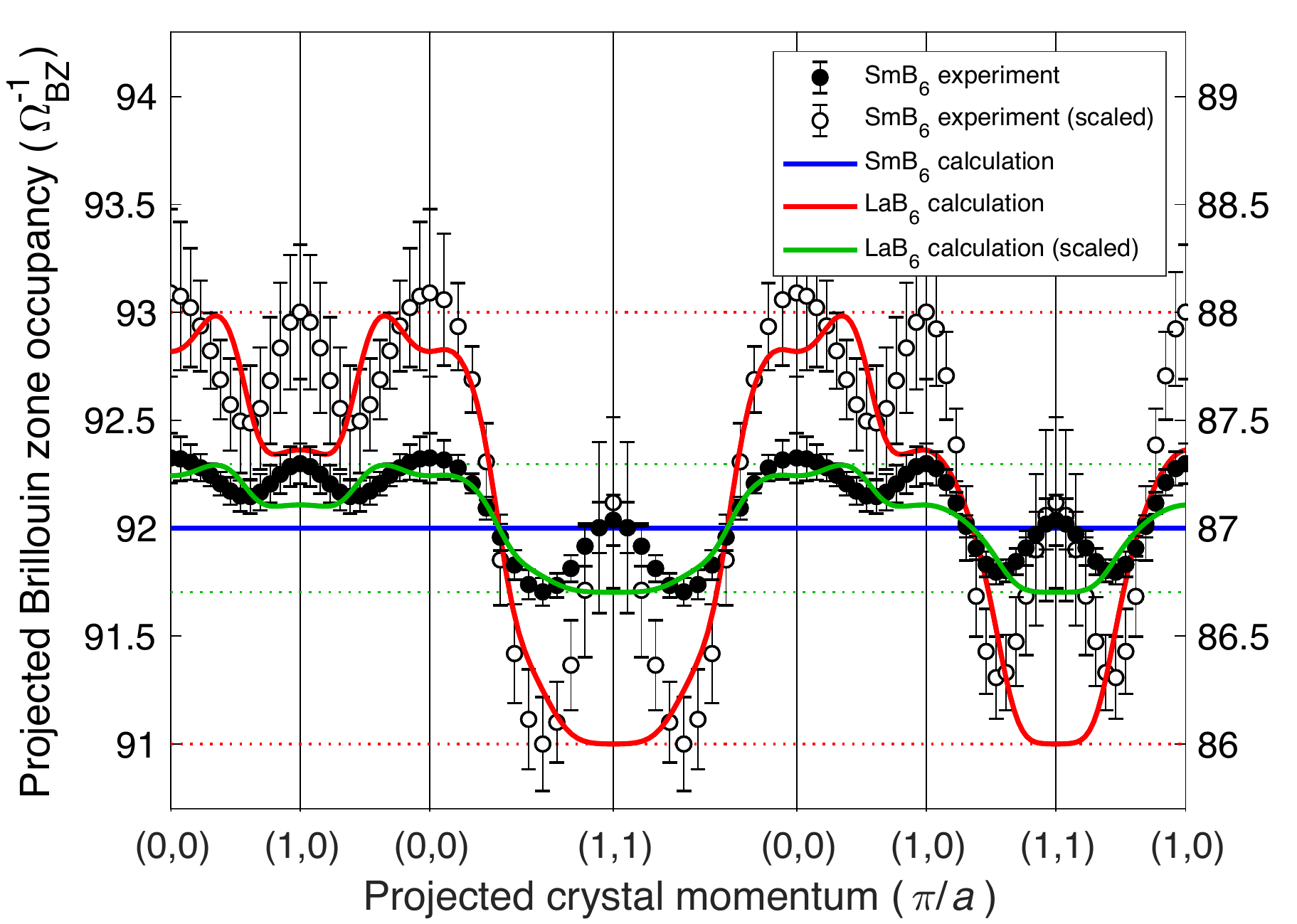}}
\caption{
{\bf Sections through the 2D projected occupation densities.}
Experimental $(001)$-plane projected $\rho_{\rm 2D}(k_{x},k_{y})$ for SmB$_{6}$ (Fig.~\ref{OccDens}{\bf a}) together with those calculated for SmB$_{6}$ and LaB$_{6}$ (Fig.~\ref{OccDens}{\bf c}) along selected 2D projected high-symmetry directions.
The horizontal dotted lines are $(N_{e}\pm1)/\Omega_{\rm BZ}$ and $(N_{e}\pm0.3)/\Omega_{\rm BZ}$, where $N_{e}=92$ and $87$ electrons per primitive cell for SmB$_{6}$ (left axis) and LaB$_{6}$ (right axis), respectively, and here $\Omega_{\rm BZ}=(2\pi/a)^{2}$ is the $(001)$-plane projected 2D Brillouin zone area.
Also shown are the experimental SmB$_{6}$ distribution scaled relative to $N_{e}/\Omega_{\rm BZ}$ to fit the LaB$_{6}$ calculation (Fig.~\ref{OccDens}{\bf b}) and {\it vice versa}.
The calculated distributions have been convoluted with a 2D Gaussian function approximating the experimental momentum resolution.
Error bars indicate statistical errors of one standard deviation, demonstrating that the observed metallic signatures are robust.
}
\label{OccPath}
\end{figure}

On back-folding the $(001)$-plane projected 2D ${\bm p}$-space distribution, $\rho_{\rm 2D}(p_{x},p_{y})$, to ${\bm k}$-space to obtain $\rho_{\rm 2D}(k_{x},k_{y})$ (see Methods),
the occupation density will only find an integer number electrons in occupied bands if no bands cross $E_{\rm F}$ along the projection direction.
For example, the calculated Fermi surface of LaB$_{6}$ viewed down the $[001]$ direction is shown in Fig.~\ref{OccDens}{\bf d} and the corresponding $(001)$-plane projected $\rho_{\rm 2D}(k_{x},k_{y})$ is shown in Fig.~\ref{OccDens}{\bf c}, with sections through $\rho_{\rm 2D}(k_{x},k_{y})$ along selected projected high-symmetry directions shown in Fig.~\ref{OccPath}.
The $(001)$-plane projection of the 3D ellipsoids gives the overall structure of $\rho_{\rm 2D}(k_{x},k_{y})$, with the four highest intensity regions close to the projected $(0,0)$-point being due to the intersection of the ellipsoids, corresponding to full occupation along the projection direction at that point.
Crucially, the underlying 3D Fermi surface is still visible in the 2D projection.

Fig.~\ref{OccDens}{\bf a} and \ref{OccPath} show the experimental $\rho_{\rm 2D}(k_{x},k_{y})$ for SmB$_{6}$.
The distribution clearly shows a remarkable departure from insulating behaviour as evidenced by the variation across the projected Brillouin zone area, as opposed to the featureless (${\bm k}$-independent) distribution expected for an insulator.
The features in $\rho_{\rm 2D}(k_{x},k_{y})$ are consistent with the Fermi surface proposed in the recent dHvA experiment \cite{Tan2015a}, strongly resembling that of metallic LaB$_{6}$ with the signature of $\big\{\frac{\pi}{a},0,0\big\}$-centred ellipsoids \cite{Biasini1997a,Harrison:1993}.
Additionally, occupied intensity is observed around the projected set of
$\big\{\frac{\pi}{a},\frac{\pi}{a}\big\}$-points, suggestive of an additional
Fermi surface component which has not been reported previously and which is not accessible by any simple rigid shift of the band structure relative to $E_{\rm F}$.
To extract the Fermi surface dimensions, the experimental $\rho_{\rm 2D}(k_{x},k_{y})$ was fitted with a 3D Fermi surface model consisting of three $\big\{\frac{\pi}{a},0,0\big\}$-centred ellipsoids and three $\big\{\frac{\pi}{a},\frac{\pi}{a},0\big\}$-centred spheres, involving just three free parameters (see Fig.~S4 and S5 in the Supplementary information).
Table~\ref{tab1} compares the experimental Fermi surface dimensions for SmB$_{6}$ extracted from the fit with those determined by other techniques
(see Fig.~S6 in the Supplementary information for a graphical comparison).

An important consideration of $\rho({\bm k})$ is the {\it intensity} of the Fermi surface signal, which is best illustrated in Fig.~\ref{OccPath}.
Intriguingly, the dynamic range of the experimental $\rho_{\rm 2D}(k_{x},k_{y})$ of SmB$_{6}$ is only about
$91.7$--$92.3$~$\Omega_{\rm BZ}^{-1}$, and integration of the dynamic range over the projected Brillouin zone
area gives only $0.3\pm0.1$ electrons which, significantly, does not match the single electron contained within the half-filled Brillouin zone volume implied by the Fermi surface dimensions (see Table~\ref{tab1}).
This remarkable situation does not correspond to conventional insulating or metallic behaviour.
To highlight the metallic steps in the 2D projected occupancy, and how this corresponds to the 3D Fermi surface, Figs.~\ref{OccDens}{\bf b} and \ref{OccPath} show $\rho_{\rm 2D}(k_{x},k_{y})$ scaled to reflect the full dynamic range of a monovalent metal ($91$--$93$~$\Omega_{\rm BZ}^{-1}$), similar to LaB$_{6}$ ($86$--$88$~$\Omega_{\rm BZ}^{-1}$),
representing what we would expect to observe were it possible to realise a conventional monovalent metallic state in SmB$_{6}$ under our experimental conditions ($T=7$~K, standard pressure and zero applied magnetic field).
Similar to LaB$_{6}$, the Fermi surface of such a metal would contain exactly one conduction electron such that the Brillouin zone volume is exactly half-filled, obeying Luttinger's theorem \cite{Luttinger1960}.
For completeness, Fig.~\ref{OccPath} also shows sections through the calculated flat distribution for insulating SmB$_{6}$, and the calculated $\rho_{\rm 2D}(k_{x},k_{y})$ for LaB$_{6}$ but with the dynamic range scaled to $86.7$--$87.3$~$\Omega_{\rm BZ}^{-1}$, such that its intensity corresponds to the $0.3$ electrons found in SmB$_{6}$.
Thus, the {\it shape} of the implied 3D Fermi surface indicates that it half-fills the Brillouin zone volume and contains one electron, but the {\it intensity} of the observed signal indicates that it contains only $0.3\pm0.1$ electrons, in apparent violation of Luttinger's theorem.

\begin{table}[b!]
\centering
\caption{Comparison of 3D (bulk) and 2D (surface) experimental Fermi surface dimensions of SmB$_{6}$. The asterisk ($^{\ast}$) indicates values determined for LaB$_{6}$.}
\begin{tabular}{l r r r}
\hline
\hline
                                            & $\big\{\frac{\pi}{a},0,0\big\}$ & $\big\{\frac{\pi}{a},0,0\big\}$ & $\big\{\frac{\pi}{a},\frac{\pi}{a},0\big\}$ \\
                                            & ellipsoids          & ellipsoids          & spheres      \\
                                            & semi-major          & semi-minor          & radius       \\
Technique                                   & radius (\AA$^{-1}$) & radius (\AA$^{-1}$) & (\AA$^{-1}$) \\
\hline
Compton\footnotemark[1] (3D)              & $0.69(8)$      & $0.38(12)$     & $0.31(15)$ \\
dHvA (3D) \cite{Tan2015a}                 & $0.69$         & $0.49$         &            \\
dHvA (2D) \cite{Li2014a}                  &                & $0.30(2)$      &            \\
ARPES (2D) \cite{Denlinger2016a}          & $0.39(1)$      & $0.27(1)$      &            \\
ARPES (2D) \cite{Jiang2013a}              & $0.42$         & $0.29$         &            \\
dHvA$^{\ast}$ (3D) \cite{Harrison:1993}   & $0.620^{\ast}$ & $0.489^{\ast}$ &            \\
\hline
\hline
\end{tabular}
\label{tab1}
\footnotetext[1]{This work.}
\end{table}

This apparent discrepancy can perhaps be understood in terms of the mixed valence Kondo behaviour of SmB$_{6}$ at low temperature and ambient pressure.
In general, Kondo insulators can be understood as half-filled (monovalent) local magnetic moment ($J\neq0$) metals in which the conduction electrons quench the local moments by forming non-magnetic Kondo singlets ($J=0$) below the Kondo temperature, $T_{\rm K}$ \cite{Coleman2007}.
This process opens a Kondo gap at $E_{\rm F}$ with the drastic consequence of appearing as though the Fermi surface has vanished (the half-filled Brillouin zone above $T_{\rm K}$ has become completely filled below $T_{\rm K}$).
The Kondo effect can be thought of as liberating a negatively charged, heavy electron which is compensated by endowing each formerly magnetic ion with an additional unit of positive charge \cite{Coleman2007}, thus ensuring Luttinger's theorem \cite{Luttinger1960} is not violated \cite{Martin1982,Oshikawa2000}.

This explanation works well in the so-called Kondo regime where the $4f$ occupation is close to integer.
In SmB$_{6}$, however, inelastic neutron scattering \cite{Alekseev1993}, Sm $L_{3}$-edge XAS \cite{Mizumaki2009a} and hard x-ray $3d$ core-level photoemission spectroscopy \cite{Utsumi2017} all report an average Sm valence close to $+2.5$ ($50\%$ Sm$^{2+}$ and $50\%$ Sm$^{3+}$) at low temperatures, placing SmB$_{6}$ ($T_{\rm K}\sim100$~K \cite{Phelan2014a}) firmly in the mixed valence regime.
Here, the proximity of the $4f$ levels to $E_{\rm F}$ leads to dynamic charge fluctuations between non-magnetic Sm$^{2+}$ ($4f^{6}$, $J_{4f}=0$) and magnetic Sm$^{3+}$ ($4f^{5}+5d^{1}$, $J_{4f}=5/2$) states.
Since the $4f^{6}$ spin-orbital singlet has no $4f$ moment, no Kondo effect occurs.
Kondo singlet formation can only be realised for the magnetic Sm$^{3+}$ state such that both $4f^{5}$ ($J_{4f}=5/2$) and $5d^{1}$ are present, in which case $4f^{5}+5d^{1}\rightarrow[4f^{5}5d^{1}]$ resulting in the Sm$^{3+}$ configuration but, crucially, with $J=0$.
Indeed, Kasuya \cite{Kasuya1994a} argued that the mixed valence state in SmB$_{6}$ at low temperature can be viewed as a coherent fluctuation between the Sm$^{2+}$ spin-orbital singlet and the Sm$^{3+}$
Kondo singlet states at each Sm site, $4f^{6}\rightleftharpoons[4f^{5}5d^{1}]$.
Importantly, both the $4f^{6}$ spin-orbital singlets and the $[4f^{5}5d^{1}]$ Kondo singlets are non-magnetic and gapped at $E_{\rm F}$ giving low temperature bulk properties that appear renormalised towards the Sm$^{2+}$ configuration (non-magnetic, electrically insulating).

To reconcile this with our observation of a remnant Fermi surface, we argue that it must be a Kondo destroyed (magnetic) $4f^{5}+5d^{1}$ state with a delocalised electron that half-fills a $5d$ band which gives the remnant Fermi surface signal.
More specifically, we believe this Kondo destroyed state exists as transient but spatially coherent metallic regions within the sample linked to the mixed valency of Sm in SmB$_{6}$.
The average energy lost by the photons in our Compton scattering experiment ($\Delta E\approx35$~keV) indicates an extremely fast interaction timescale, $\tau_{\rm CS}=h/\Delta E\sim10^{-19}$~s, compared with other techniques such that we, essentially, measure a series of instantaneous snapshots of the fluctuating valence in which each Sm ion within the scattering volume is in the magnetic Sm$^{3+}$ state for about $30\%$ of the time.
The $50\%$ Sm$^{2+}$ population is solely from the non-magnetic $4f^{6}$ state, while the $50\%$ Sm$^{3+}$ population is split between non-magnetic, Kondo insulating $[4f^{5}5d^{1}]$ states and magnetic, metallic $4f^{5}+5d^{1}$ states with average populations of about $20\%$ and $30\%$, respectively.

Evidence for such a state comes from a variety of techniques.
Inelastic neutron scattering observes both $J_{4f}=0\rightarrow1$ and $J_{4f}=5/2\rightarrow7/2$ spin-flip excitations \cite{Alekseev1993}, with the latter excitation only attributable to the magnetic (metallic) Sm$^{3+}$ state,
and $\mu$SR measurements find magnetic fluctuations that are homogeneous, dynamic and spatially coherent up to $\xi\lesssim100$~nm \cite{Biswas2014a,Biswas2017a}, rather than existing as inhomogeneous spatially static regions.
Remarkably, application of relatively modest hydrostatic pressure ($4$--$10$~GPa) is sufficient to completely destroy the Kondo gap in SmB$_{6}$ \cite{Barla2005a,Derr2008a,Butch16,Zhou2017,chen:18}.
XAS has tracked the Sm$^{2+}$ and Sm$^{3+}$ populations as a function of pressure through the metallic transition and found that
the Sm$^{3+}$ state dominates above $10$~GPa \cite{Butch16,Zhou2017}.
Furthermore, nuclear forward scattering has measured a magnetic fraction of $30\%$ at low temperature ($3$~K) and ambient pressure \cite{Barla2005a}, exactly matching the metallic fraction extracted from Compton scattering, which saturates to $100\%$ at $10$~GPa \cite{Derr2008a}.
In our picture of transient metallic fluctuations, we envisage that the transition under hydrostatic pressure from a mixed valence Kondo insulator to a local moment metal begins to happen when the percolation threshold is reached such that bulk electrically conducting Sm$^{3+}$ channels start to completely extend across the macroscopic sample dimensions.

In conclusion, the variation in occupation density observed by Compton scattering matches that of infinitely sharp steps (within the experimental momentum resolution), strongly indicative of the Fermi level lying within a band, the signature of a metallic state.
The observed reduction in Fermi surface signal intensity is inconsistent with the filling
of the Brillouin zone and indicates that the bulk is not completely metallic or insulating
in the conventional sense.
Once again, SmB$_{6}$ is seen to display an {\it unconventional} Fermi surface in the bulk.
\\

\noindent
{\large {\bf Methods}}
\\
{\small
\noindent
{\bf Crystal growth.}
The SmB$_{6}$ single crystal used for this study was grown at the University of Warwick, UK, using the floating zone technique \cite{Hatnean2013a}.
\\

\noindent
{\bf Compton scattering.}
In Compton scattering, monochromatic synchrotron photons are scattered incoherently from electrons
in the target material.
The measured distribution in energy of scattered photons is Doppler broadened as a consequence of the electron momentum component, $p_{z}$, parallel to the scattering vector.
The broadened distribution is then proportional to the Compton profile, $J(p_{z})$.
Compton scattering was performed on beamline BL08W, SPring-8, using $115$~keV monochromatic synchrotron x-rays.
Measurements were performed at the base temperature of the cryostat, $T=7$~K ($k_{\rm B}T=0.6$~meV), with standard pressure and zero
applied magnetic field.
Six different directions within the $(001)$-plane were measured, spaced equally between (and including)
the $[100]$ and $[110]$ directions.
The full-width-half-maximum momentum resolution was $0.11$~a.u. at the Compton peak.
\\

\noindent
{\bf Back-folding from ${\bm p}$-space to ${\bm k}$-space.}
To transform the measured ${\bm p}$-space distributions to ${\bm k}$-space, the back-folding process
follows the method of Lock, Crisp and West \cite{Lock1973a}.
Recalling that ${\bm p}={\bm k}+{\bm G}$, where ${\bm G}$ is any vector of the reciprocal lattice,
in 3D we can write $\rho({\bm k})=\sum_{\bm G}\rho({\bm p}+{\bm G})$, where the (infinite) sum is over
all ${\bm G}$-vectors.
In a 2D projection, $\rho_{\rm 2D}(k_{x},k_{y})=\sum_{(G_{x},G_{y})}\rho_{\rm 2D}(p_{x}+G_{x},p_{y}+G_{y})=\int_{\Omega_{\rm BZ}}\rho({\bm k})~\text{d}k_{z}$, where $(G_{x},G_{y})$ are the 2D projected reciprocal lattice vectors and $\Omega_{\rm BZ}$ is the 1D projected Brillouin zone length.
In a 1D projection, $\rho_{\rm 1D}(k_{z})=\sum_{G_{z}}J(p_{z}+G_{z})=\iint_{\Omega_{\rm BZ}}\rho({\bm k})~\text{d}k_{x}\text{d}k_{y}$, where $G_{z}$ are the 1D projected reciprocal lattice vectors and $\Omega_{\rm BZ}$ is the 2D projected Brillouin zone area.
As $\rho({\bm p})$ is the probability density of electrons in momentum space, once integrated over all momenta, it must equal the number of electrons per primitive cell, $N_{e}$.
The same must be true even when measured in projection and subsequently back-folded into the first Brillouin zone:
$N_{e}\equiv\iiint_{\Omega_{\rm BZ}}\rho({\bm k})~\text{d}^{3}{\bm k}\equiv\iiint\rho({\bm p})~\text{d}^{3}{\bm p}$, where $\Omega_{\rm BZ}=(2\pi/a)^{3}$ is the 3D Brillouin zone volume.
Thus, $\rho_{\rm 2D}(k_{x},k_{y})$ can also be normalised such that it contains $N_{e}$ electrons per primitive cell.
\\

\noindent
{\bf Counting electrons.}
The obtained $\rho({\bm k})$ distribution can be directly related to the electron energy bands,
$E_{i}({\bm k})$, by writing $\rho({\bm k})=\sum_{i}n_{i}({\bm k})$, where $n_{i}({\bm k})$ is
the occupation density of the $i^{\rm th}$ band at a given ${\bm k}$-point.
In a one-electron framework, $n_{i}({\bm k})$ is straightforwardly related to $E_{i}({\bm k})$
through the (appropriately normalised) Fermi-Dirac distribution,
\begin{equation}
n_{i}({\bm k})=\frac{2}{\Omega_{\rm BZ}}\left[\frac{1}{\exp{\left(\frac{E_{i}({\bm k})-E_{\rm F}}{k_{\rm B}T}\right)+1}}\right],
\label{FD}
\end{equation}
where $\Omega_{\rm BZ}=(2\pi/a)^{3}$ is the 3D Brillouin zone volume, $a=7.815$~a.u. is the cubic lattice constant,
$E_{\rm F}$ is the Fermi level, $T=7$~K is the temperature, $k_{\rm B}$ is Boltzmann's constant and the factor $2$ is for spin-degeneracy.
Completely filled bands, $E_{i}({\bm k})<E_{\rm F}$, have $n_{i}({\bm k})=2/\Omega_{\rm BZ}$
and contain $\iiint_{\Omega_{\rm BZ}}n_{i}({\bm k})~\text{d}^{3}{\bm k}\equiv2$ electrons each, contributing a flat, featureless background
to $\rho({\bm k})$, while completely empty bands, $E_{i}({\bm k})>E_{\rm F}$, have $n_{i}({\bm k})=0$
and obviously contribute nothing.
It is only the partially filled bands, $E_{j}({\bm k})$, which cross $E_{\rm F}$ at ${\bm k}_{\rm F}$
(defining the Fermi surface) that will abruptly change from being occupied to unoccupied,
meaning that $n_{j}({\bm k})$ will exhibit an integer step (in $2/\Omega_{\rm BZ}$ units) at
${\bm k}={\bm k}_{\rm F}$, thus giving rise to any variation in $\rho({\bm k})$ above the
constant background.
Obviously, a partially filled band contains $0<\iiint_{\Omega_{\rm BZ}}n_{j}({\bm k})~\text{d}^{3}{\bm k}<2$ electrons.
\\

\noindent
{\bf Reciprocal form factor.}
The reciprocal form factor, $B({\bm r})$, is the autocorrelation of the ${\bm r}$-space wavefunction, $\psi({\bm r})$, and is equivalently given by the 3D Fourier transform of the ${\bm p}$-space electron momentum density, $B({\bm r})\equiv\iiint\psi^{\ast}({\bm r}')\psi({\bm r}-{\bm r}')~\text{d}^{3}{\bm r}'\equiv\iiint\rho({\bm p})\text{e}^{\text{i}{\bm p}\cdot{\bm r}}~\text{d}^{3}{\bm p}$.
Importantly, $B({\bm r})$ can be determined along a line in ${\bm r}$-space, say ${\bm r}=z$,
by taking the 1D Fourier transform of the Compton profile,
$B(z)=\iiint\rho({\bm p})\text{e}^{\text{i}p_{z}z}~\text{d}^{3}{\bm p}=\int J(p_{z})\text{e}^{\text{i}p_{z}z}~\text{d}{p_{z}}$.
Since $\rho({\bm k})=\sum_{\bm G}\rho({\bm p}+{\bm G})$ is constant (${\bm k}$-independent) for an insulator in a one-electron framework, taking the Fourier transform leads directly to the result $B(\{{\bm R}\})=0$ at non-zero integer multiples of the lattice vectors, ${\bm R}$.
Thus, $B(\{{\bm R}\})\neq0$ means $\rho({\bm k})$ is not ${\bm k}$-independent, indicative of metallic behaviour.
\\

\noindent
{\bf Electronic structure calculations.}
Density functional theory (DFT) electronic structure calculations were performed for SmB$_{6}$ and LaB$_{6}$
with the all-electron APW+l.o. code, \textsc{Elk} \cite{ELK}, using the experimental SmB$_{6}$ crystal
structure \cite{Hatnean2013a} and a $[{\rm Xe}]$ core electron configuration for Sm and La.
Convergence was realised with a $32\times32\times32$ ${\bm k}$-point grid, giving $969$ distinct
${\bm k}$-points in the irreducible Brillouin zone, with an interstitial planewave cutoff
determined by $R_{\rm mt}|({\bm G}+{\bm k})_{\rm max}|=8.5$, where $R_{\rm mt}=1.74$~a.u. is the average
muffin-tin radius (the muffin-tin radii for Sm, La and B were $2.80$~a.u., $2.80$~a.u. and
$1.56$~a.u., respectively).
Since Sm and La have relatively high atomic numbers, spin-orbit coupling was included in the calculations
by adding a term of the form ${\bm S}\cdot{\bm L}$ (where ${\bm S}$ is the spin vector
and ${\bm L}$ is the orbital angular momentum vector) to the second variational Hamiltonian.
These calculations predict the ground state of SmB$_{6}$ to be insulating with a small indirect hybridisation
gap of $\Delta\approx3$~meV ($\Delta/k_{\rm B}\approx35$~K), and the ground state of LaB$_{6}$ to be metallic, and are generally in good agreement with previous calculations \cite{antonov:02b,Hasegawa1977}.
The Fermi surface of LaB$_{6}$ was evaluated on a $64\times64\times64$ ${\bm k}$-point grid.
Valence electron Compton profiles were calculated using the method of Ernsting {\it et al.}
\cite{Ernsting2014a} out to $|{\bm p}_{\rm max}|=16$~a.u. for SmB$_{6}$ and $|{\bm p}_{\rm max}|=12$~a.u. for LaB$_{6}$ (see Fig.~S7 in the Supplementary information for a comparison of experimental and calculated SmB$_{6}$ valence electron Compton profiles for the different crystallographic directions).
}

\noindent
{\large {\bf Acknowledgements}}
\\
{\small
\noindent
The Compton scattering experiments were performed with the approval of the Japan Synchrotron Radiation Research Institute (JASRI), SPring-8, proposal numbers 2016A1640 and 2019A1460.
We thank Y. Sakurai (SPring-8) for sharing the Si experimental Compton scattering data.
This work is supported by UK EPSRC, grant numbers EP/L015544/1 and EP/S016465/1.
The crystal growth at the University of Warwick is supported by UK EPSRC, grant numbers EP/M028771/1 and EP/T005963/1.
}
\\

\noindent
{\large {\bf Author contributions}}
\\
{\small
\noindent
T.E.M., S.R.G. and S.B.D. had the initial idea for the experiment.
M.C.H. and G.B. fabricated the sample.
T.E.M., D.B., H.C.R., D.O., J.A.D., J.W.T., S.R.G. and S.B.D. performed the experiments.
T.E.M., D.B. and S.B.D. analysed the data.
T.E.M., D.B., J.L., S.R.G. and S.B.D. interpreted the results.
T.E.M., D.B., S.R.G. and S.B.D. wrote the paper with contributions from all authors.
}
\\

\noindent
{\large {\bf Additional information}}
\\
{\small
\noindent
Supplementary information is available.
}
\\

\noindent
{\large {\bf Competing interests}}
\\
{\small
\noindent
The authors declare no competing interests.
}

\end{document}


\title{Supplementary information: Direct measurement of a remnant Fermi surface in $\mathbf{SmB_{6}}$}

\date{\today}

\maketitle

\beginsupplement

\begin{figure}
\centerline{\includegraphics[width=1.0\linewidth]{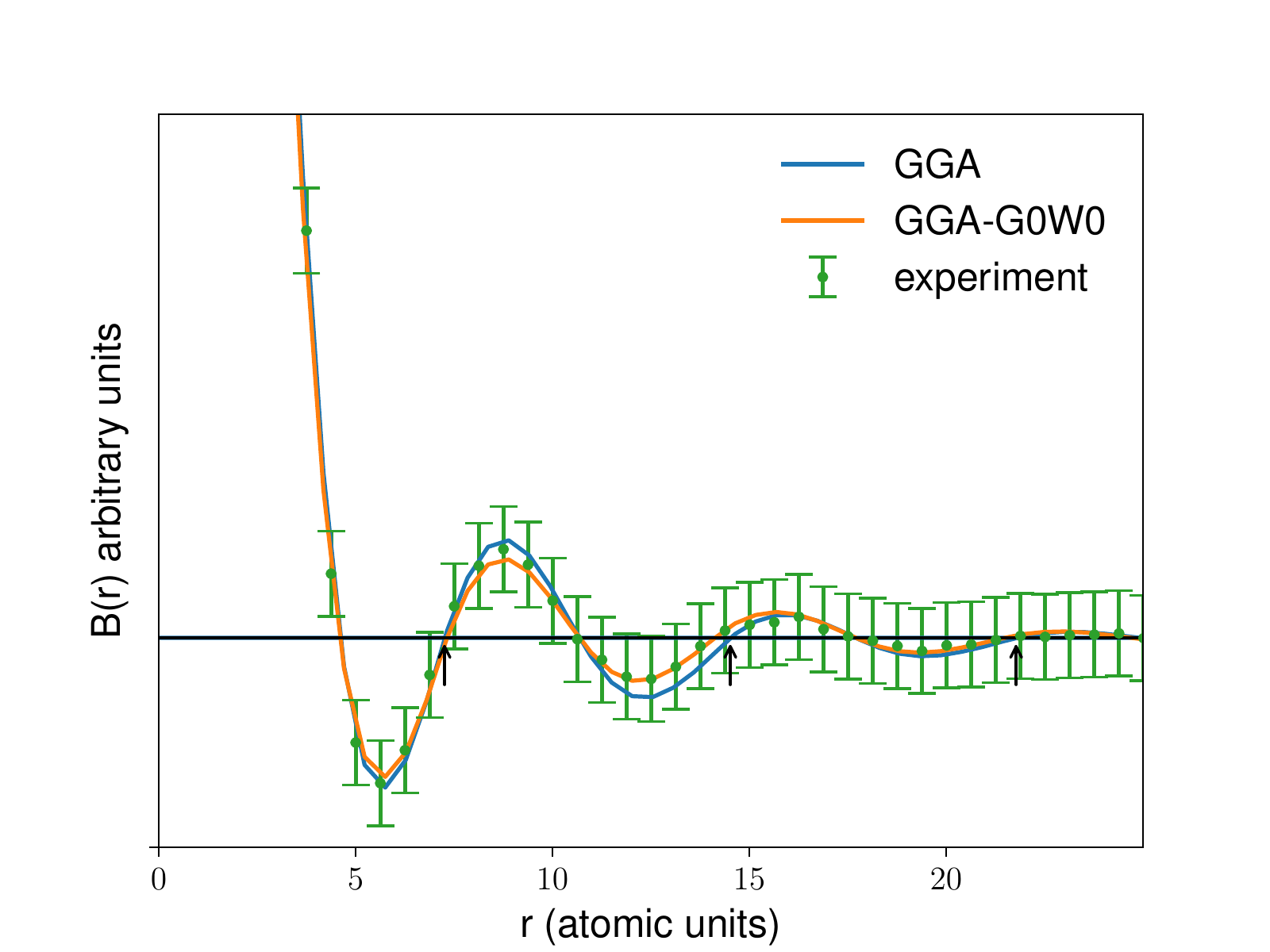}}
\caption{
Reciprocal form factor, $B(r)$, for the weakly correlated (band) insulator, Si, determined from the experimental
$[110]$ Compton profile recorded at room temperature, together with those calculated using the generalised gradient approximation (GGA) and GGA-$G_{0}W_{0}$ treatments of exchange and correlation.
Error bars indicate statistical errors of one standard deviation.
The vertical arrows indicate integer multiples of the $[110]$ lattice constant for which $B(r)$ is zero for an insulator in a one-electron framework.
}
\end{figure}

\begin{figure}
\centerline{\includegraphics[width=1.0\linewidth]{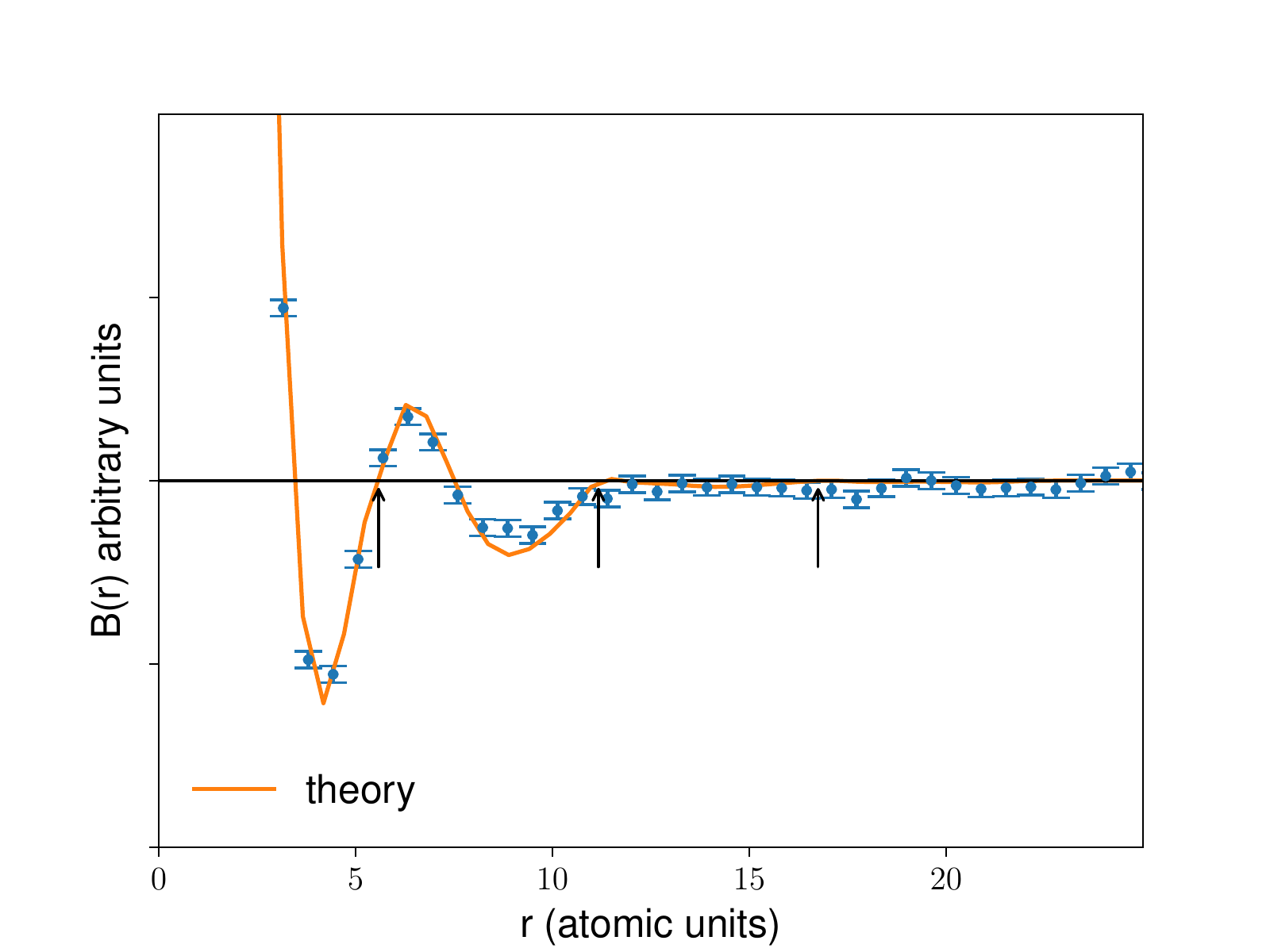}}
\caption{
Reciprocal form factor, $B(r)$, for the strongly correlated (Mott) insulator, NiO, determined from the experimental
$[110]$ Compton profile recorded at room temperature, together with that calculated for the antiferromagnetic phase using the meta-GGA SCAN functional for exchange and correlation which produces an insulating (gapped) ground state.
Error bars indicate statistical errors of one standard deviation.
The vertical arrows indicate integer multiples of the $[110]$ lattice constant for which $B(r)$ is zero for an insulator in a one-electron framework.
Clearly, the experiment is very close to zero at these points compared with the situation for SmB$_{6}$ (see Fig.~1{\bf a} of the main text).
}
\end{figure}

\begin{figure}
\centerline{\includegraphics[width=1.0\linewidth]{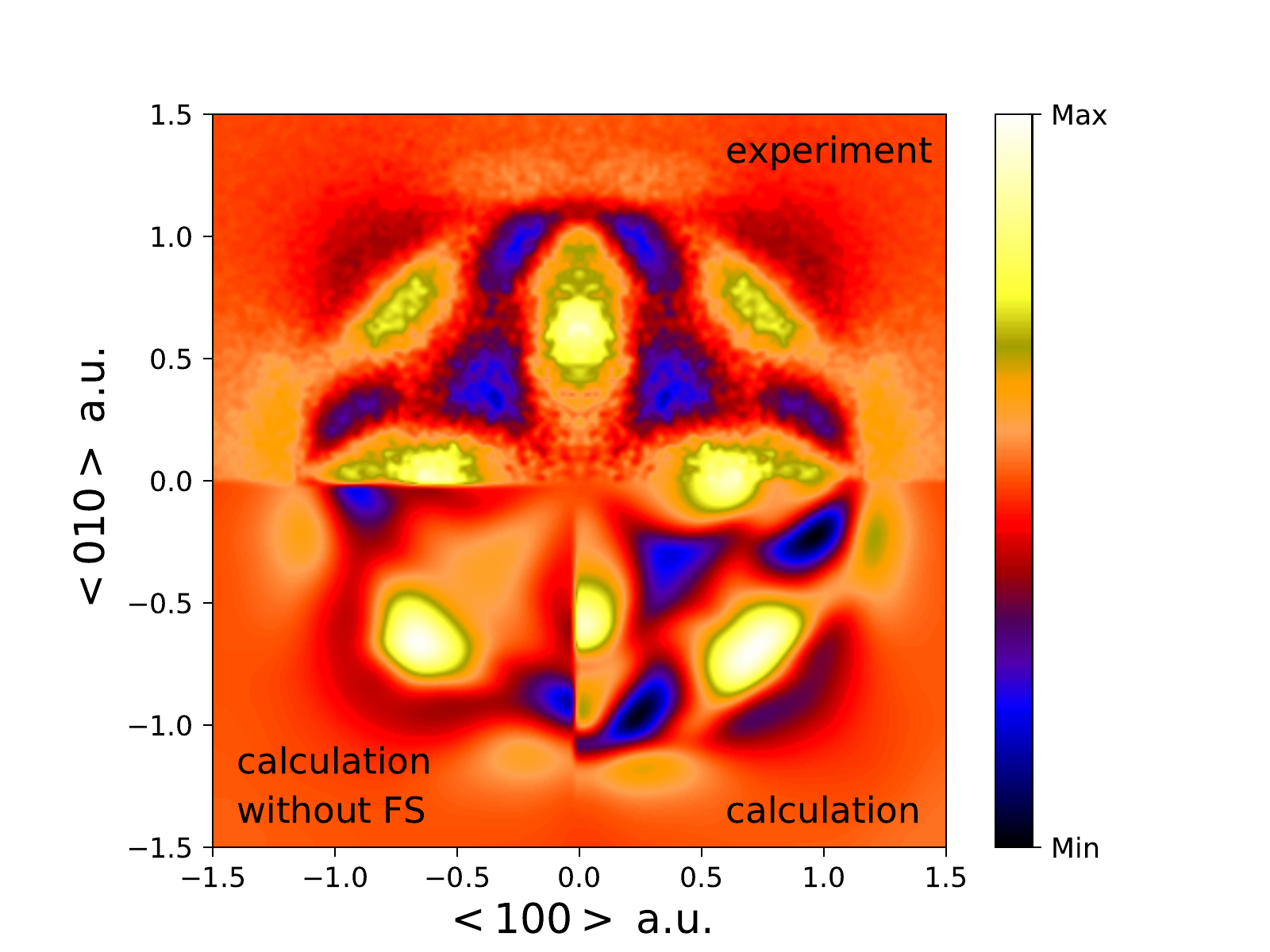}}
\caption{
Radial anisotropy of the experimental (upper half) and calculated (lower right quarter)
2D projected two-photon momentum density, $\rho_{\rm 2D}^{2\gamma}(p_{x},p_{y})$,
from 2D angular correlation of (electron-positron) annihilation radiation (2D-ACAR) for LaB$_{6}$.
The lower left quarter shows the radial anisotropy of $\rho_{\rm 2D}^{2\gamma}(p_{x},p_{y})$
calculated for only the bands which do not cross the Fermi level, $E_{\rm F}$, meaning that the
Fermi surface contribution has been completely removed from this distribution.
Clearly, the ellipsoidal features at the projected set of $\left\{\frac{\pi}{a},0\right\}$-points
are unambiguously from the Fermi surface.
Similar features are observed in the SmB$_{6}$ experiment (see main text).
The experimental data is from Biasini {\it et al.} (see Ref.~28 of the main text).
The calculated distributions have been convoluted with a 2D Gaussian function approximating the experimental momentum resolution.
}
\end{figure}

\begin{figure}
\centerline{\includegraphics[width=1.0\linewidth]{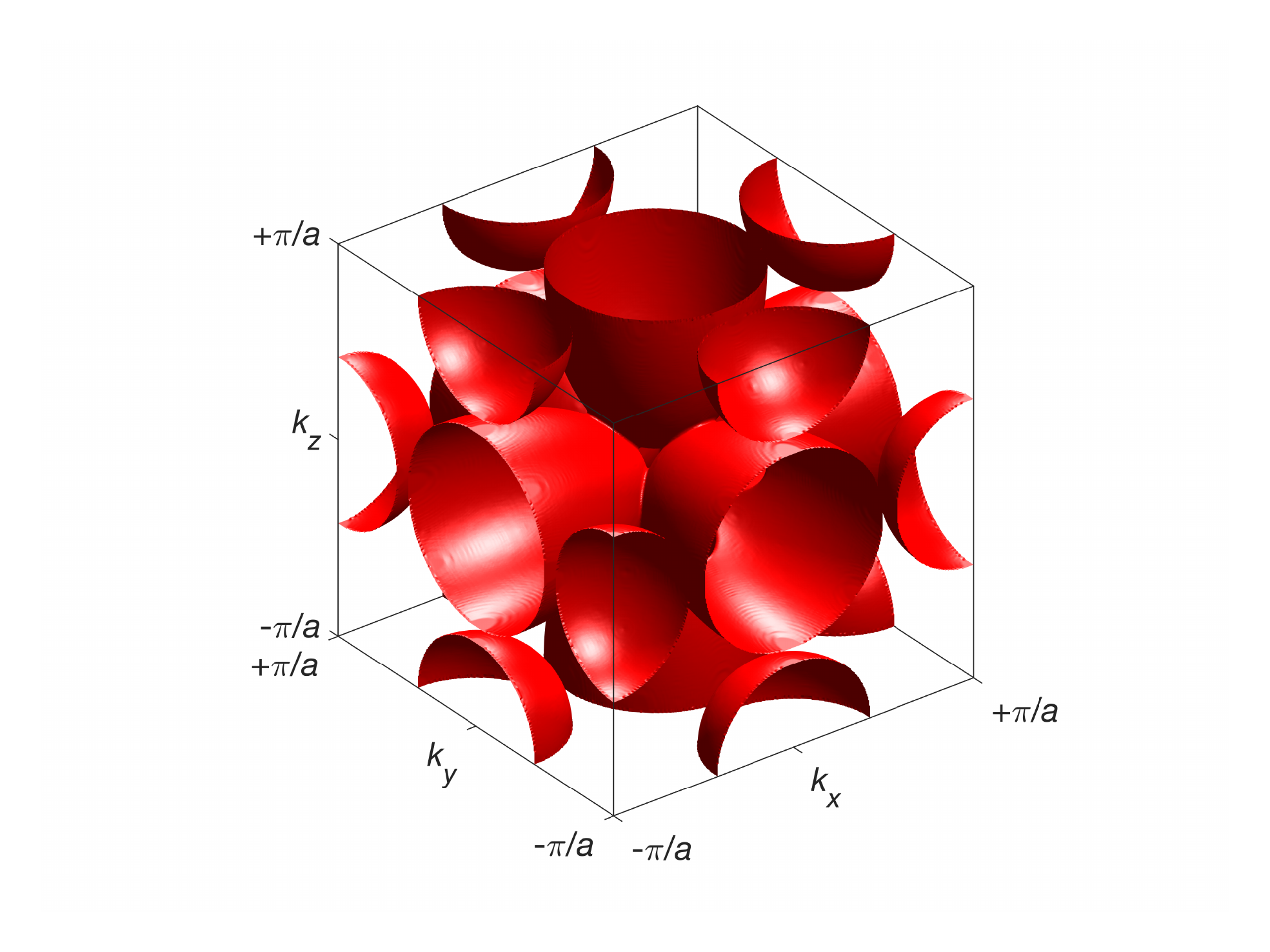}}
\caption{
Best fit Fermi surface model for SmB$_{6}$, consisting of three intersecting ellipsoids centred
at the set of $\left\{\frac{\pi}{a},0,0\right\}$-points and three spheres centred at the
set of $\left\{\frac{\pi}{a},\frac{\pi}{a},0\right\}$-points.
This was determined by varying three free parameters, namely the semi-major and semi-minor
axes of the ellipsoid and the radius of the sphere (given in Table~1 of the main text),
and minimising the square of the difference between the scaled (see Fig.~2{\bf b} and Fig.~3 of the main text) experimental
$\rho_{\rm 2D}(k_{x},k_{y})$ for SmB$_{6}$ and the same quantity determined
from the 3D Fermi surface model convoluted with a 2D Gaussian function approximating the experimental momentum resolution (see Fig.~S5 and the upper right quarter of Fig.~S6).
}
\end{figure}

\begin{figure}
\centerline{\includegraphics[width=1.0\linewidth]{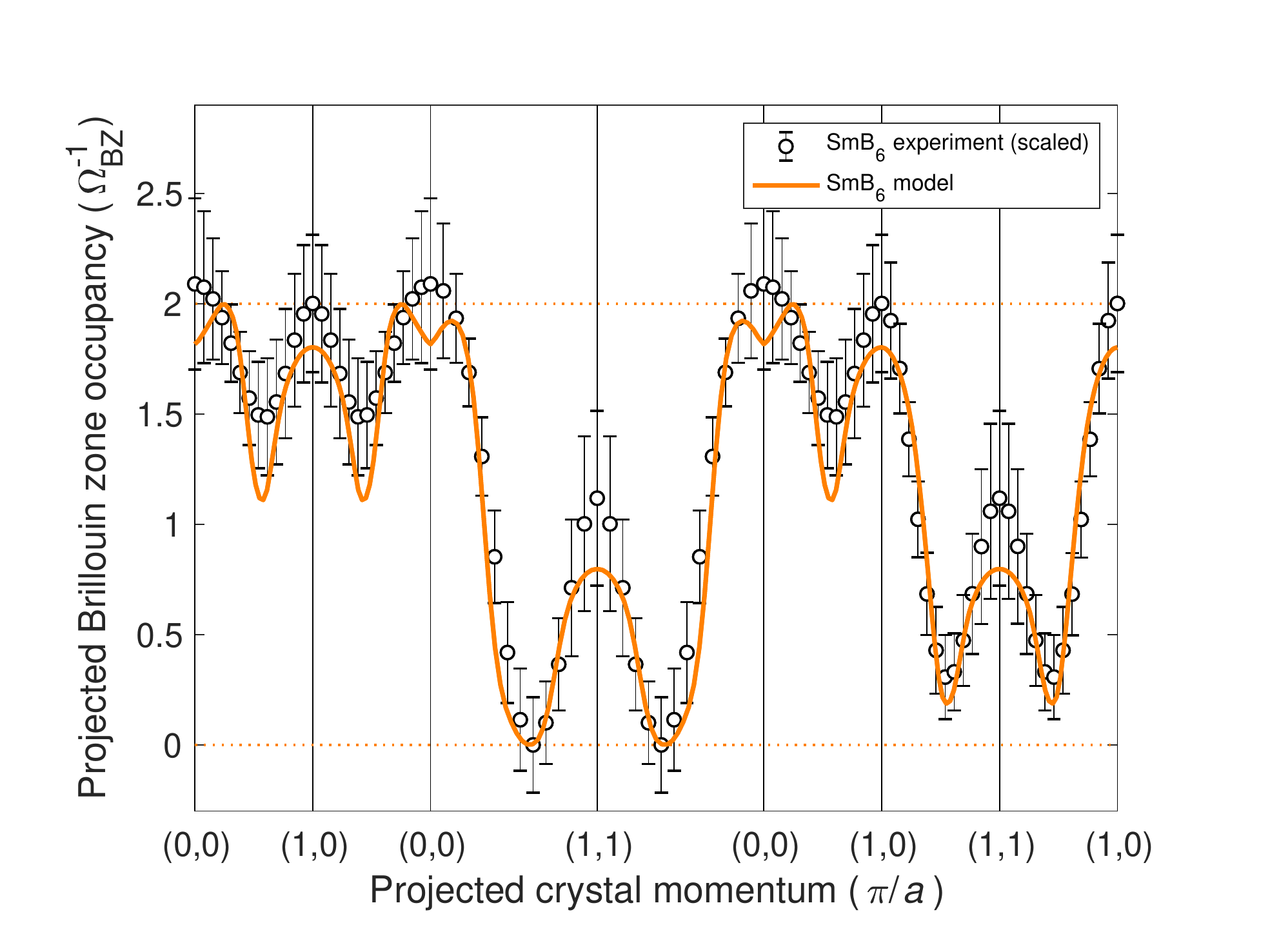}}
\caption{
Sections through the 2D projected occupation density determined from the model Fermi surface (see Fig.~S4) compared with sections through the scaled (see Fig.~2{\bf b} and Fig.~3 of the main text) experimental $(001)$-plane projected $\rho_{\rm 2D}(k_{x},k_{y})$ for SmB$_{6}$ along selected 2D projected high-symmetry directions.
The 2D distribution (shown in the upper right quarter of Fig.~S6) determined from the Fermi surface model (see Fig.~S4) was convoluted with a 2D Gaussian function approximating the experimental momentum resolution.
Error bars indicate statistical errors of one standard deviation.
The horizontal dotted lines are $(N_{e}\pm1)/\Omega_{\rm BZ}$ where, for ease of comparison, $N_{e}=1$ conduction electron per primitive cell and, here, $\Omega_{\rm BZ}=(2\pi/a)^{2}$ is the $(001)$-plane projected 2D Brillouin zone area.
}
\end{figure}

\begin{figure}
\centerline{\includegraphics[width=1.0\linewidth]{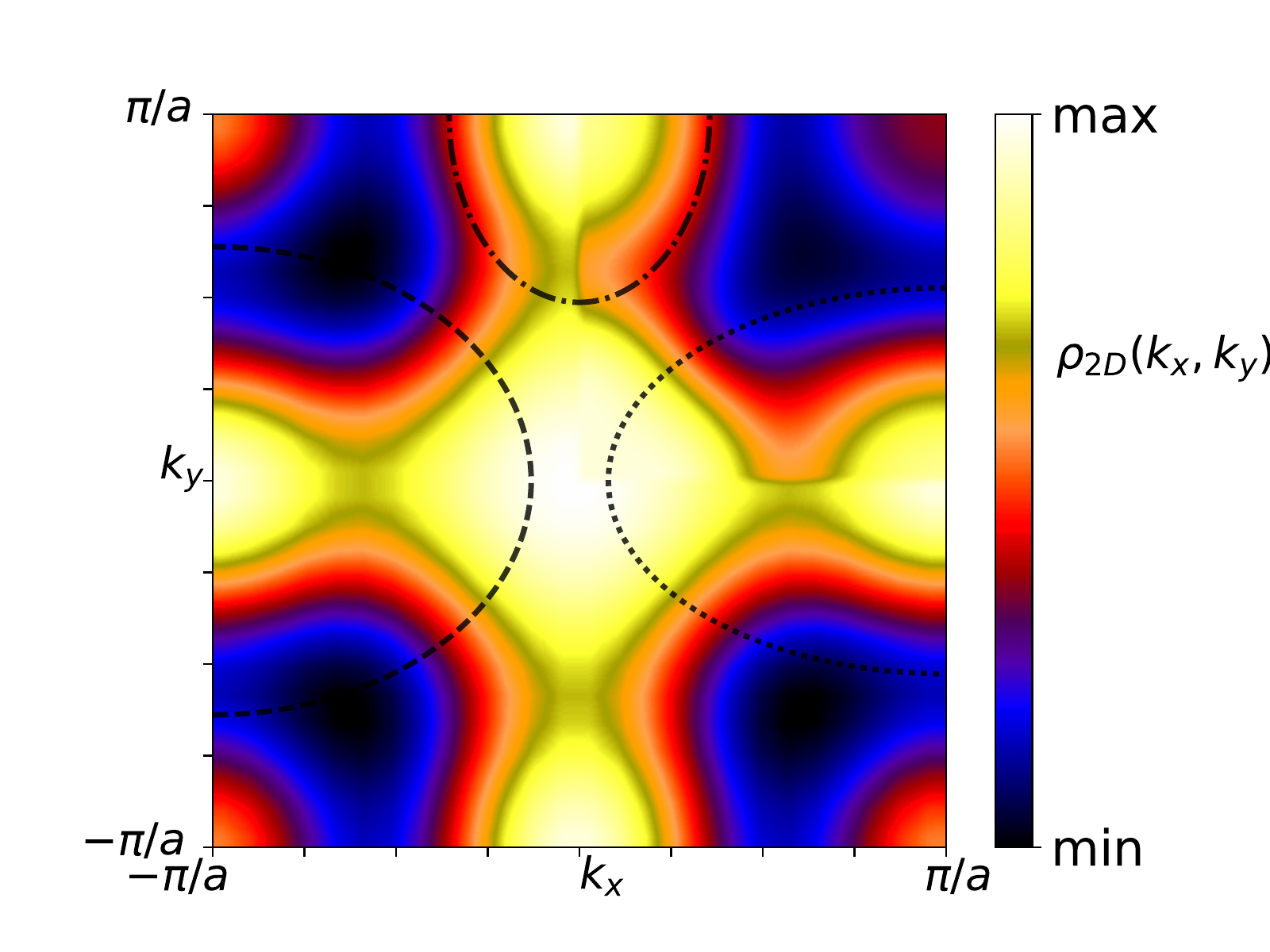}}
\caption{
Experimental 2D projected occupation density, $\rho_{\rm 2D}(k_{x},k_{y})$,
for SmB$_{6}$.
The upper right quarter shows the comparable distribution determined from the Fermi surface model (see Fig.~S4 and S5).
Ellipses demonstrating the dimensions of the ellipsoidal component of the remnant Fermi surface are plotted for the Fermi surface model in
Fig.~S4 and S5 (dotted line, right).
Also shown are the small ellipse from ARPES (dot-dashed line, top) and the ellipse for LaB$_{6}$ (dashed line, left, from Ref.~29 of the main). See Table~1 of the main text for the quantitative dimensions.
}
\end{figure}

\begin{figure}
\centerline{\includegraphics[width=1.0\linewidth]{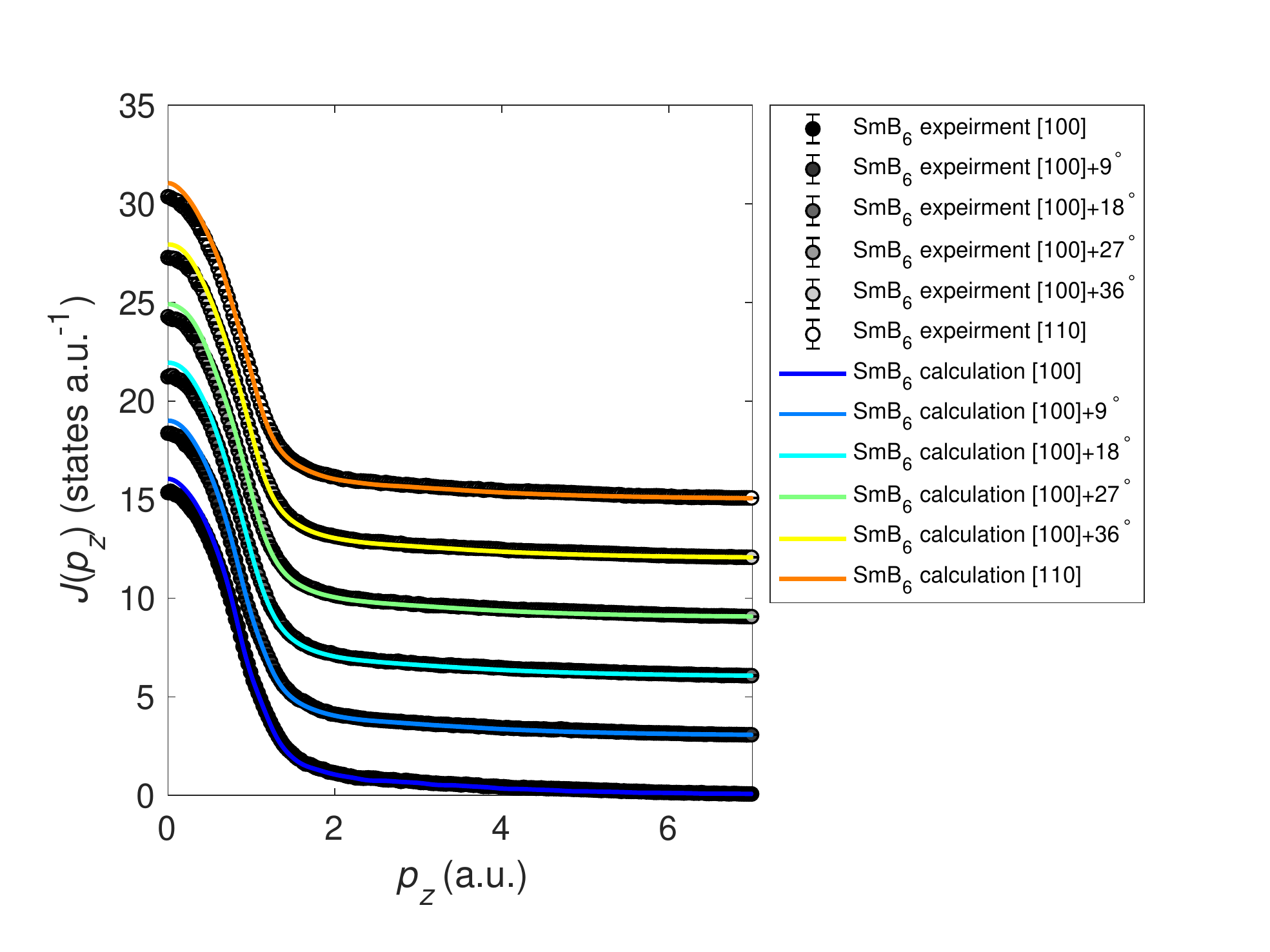}}
\caption{
Experimental and calculated valence electron Compton profiles, $J(p_{z})$, for SmB$_{6}$
(the isotropic [Xe] core electron configuration has been removed).
The calculated profiles have been convoluted with a 1D Gaussian function approximating the experimental momentum resolution.
The angles in the legend are from the $[100]$ to $[110]$ directions in the $(001)$-plane
($[110]=[100]+45^{\circ}$).
The profiles have been vertically offset from each other by $3$~states a.u.$^{-1}$ for clarity.
}
\end{figure}